\documentclass[preprints,article,accept,pdftex,moreauthors]{Definitions/mdpi}

\firstpage{1} 
\pubvolume{1}
\issuenum{1}
\articlenumber{0}
\pubyear{2025}
\copyrightyear{2025}
\externaleditor{Firstname Lastname} 
\datereceived{ } 
\daterevised{ } 
\dateaccepted{ } 
\datepublished{ } 
\hreflink{https://doi.org/} 

\Title{An Updated Repository of Sub-mJy Extragalactic Source-Count Measurements in the Radio Domain}

\TitleCitation{An Updated Repository of Sub-mJy Extragalactic Source-Count Measurements in the Radio Domain}

\Author{Vincenzo 
 Galluzzi $^{1,2,}$*\orcidA{}, Meriem Behiri $^{3}$\orcidB{}, Marika Giulietti $^{1}$\orcidC{} and Andrea Lapi $^{1,3,4,5}$\orcidD{}}

\AuthorNames{Vincenzo Galluzzi, Meriem Behiri, Marika Giulietti, and Andrea Lapi}

\AuthorCitation{Galluzzi V. ;  Behiri, M. ; Giulietti, M.; Lapi, A.}

\address{%
$^{1}$ \quad INAF---Istituto di Radioastronomia, 
Via Gobetti 101, 40129 Bologna, Italy; marika.giulietti@inaf.it (M.G.);  lapi@sissa.it (A.L.) 
\\
$^{2}$ \quad INAF---Osservatorio Astronomico di Trieste \& Italian Centre for Astronomical Archives, Via Tiepolo, 11, 34131 Trieste, Italy\\ 
$^{3}$ \quad Scuola Internazionale Superiore di Studi Avanzati, Via Bonomea 265, 34136 Trieste, Italy;  mbehiri@sissa.it\\
$^{4}$ \quad Institute for Fundamental Physics of the Universe (IFPU), Via Beirut 2, 34014 Trieste, Italy\\
$^{5}$ \quad INFN-Sezione di Trieste, via Valerio 2, 34127 Trieste, Italy}

\corres{Correspondence: vincenzo.galluzzi@inaf.it}

\abstract{We present an updated repository of sub-mJy extragalactic radio source counts between $150$ MHz and $10$ GHz, incorporating recent advances in radio surveys and observational techniques. By compiling and refining previous datasets, we provide a comprehensive catalog that enhances the understanding of faint radio-source populations, including Dusty Star-Forming Galaxies (DSFGs) and Radio-Quiet Active Galactic Nuclei (RQAGNs), from intermediate to high redshifts. Our analysis accounts for observational biases, such as resolution effects and Eddington bias, ensuring improved accuracy in flux-density estimations. We also discuss the implications of new-generation radio telescopes, such as the Square-Kilometer Array Observatory (\texttt{SKAO}) and its precursors and pathfinders, to further resolve these populations. Our collection 
 contributes to constraining evolutionary models of radio sources, highlighting the increasing role of polarization studies in distinguishing different classes. This work serves as a key reference for future deep radio surveys targeting the faintest end of the extragalactic radio sky.}

\keyword{extragalactic 
 radio sources; sub-mJy surveys; radio-quiet AGNs; star-forming galaxies; source counts; polarization; \texttt{SKA}}

\begin{document}

\section{Introduction}
\label{sec:intro}

Source counts of extragalactic radio sources provide information about the underlying cosmology \citep{Ryle1955,ryle1961,longair1966,longair1974} and are an essential tool for defining the different radio-source populations (see \cite{condon2012} and reference therein). 
In particular, \cite{dezotti2010} collected a comprehensive review of the contribution to the millimeter surveys available at the time coming from the radio source counts. The review was completed by the release of a collection of available source counts, also exploited in \cite{massardi2011} to define an evolutionary model for Active Galactic Nuclei (AGNs), updated later by \cite{Bonato2017} and recently integrated in the Tiered Radio Extragalactic Continuum Simulation \texttt{T-RECS}; \cite{Bonaldi2019,Bonaldi2023}). 

The expected upgrade refurbishment of existing interferometric facilities (e.g., \texttt{ATCA}, \url{https://www.atnf.csiro.au/projects/instrumentation/bigcat/}, accessed on 28 March 2025), 
 \texttt{VLA} \cite{wilner2024,selina23}, 
 \texttt{GMRT} {\cite{gupta17}}) and the emergence of several crucial actors paving the way to the \texttt{SKA} (e.g., \texttt{A\texttt{SKA}P} \textls[-15], {{{\url{https://www.csiro.au/en/about/facilities-collections/ATNF/ASKAP-radio-telescope}}}), }\linebreak  
 \texttt{MeerKAT}, {{\url{https://www.sarao.ac.za/science/meerkat/about-meerkat/}}}, \texttt{LOFAR}, \linebreak  {{\url{https://lofar-surveys.org/}}}) has substantially improved both the efficacy and accuracy of calibration, imaging, and source detection techniques, therefore improving the definition of source counts, in particular toward the faintest end. 

While local Active Galactic Nuclei (AGNs) dominate the bright end of the radio counts, such a fainter sub-mJy source population is a combination of different objects, most noticeably including Dusty Star-Forming Galaxies (DSFGs) and Radio-Quiet AGNs (RQAGNs) from intermediate to high redshift ($z\gtrsim 2$), the relative amount of which is still under debate.  

On the one hand, the radio AGN spectral behavior is usually characterized by a superposition of synchrotron components, tending to a flatter spectrum when multiple compact jetted synchrotron emissions overlap. Considering relativistic bulk motions that might be present in components closer to the central engine (whose synchrotron peak is at higher frequencies as they are closer), Doppler boosting can significantly enhance fluxes of sources and contribute to their flat-spectrum appearance wherever the line of sight is close to the jet axis. Instead, a steeper spectrum behavior is generally associated with cases where self-absorption is less effective, e.g., in more extended objects or as a result of the environment surrounding the central AGN. Peaking spectra where one or more components can be identified are generally associated with newly jetted components, e.g., as a result of a re-activation of the AGN \cite{retriggered}, or of an ejection axis particularly close to the line of sight, or of a prominent flaring activity (e.g., \cite{slob,callingham}). 

Moreover, the synchrotron signal can be intrinsically linearly polarized up to $\sim$70\%. However, observed polarization fractions are at a low $\%$ level, as source geometry and complexities in the magneto-ionic medium, together with poor instrumental spatial or spectral resolutions, induce substantial depolarization effects. Nevertheless, deeper surveys performed with newer facilities (such as \texttt{SKA} precursors and pathfinders) characterized by improved survey speed, deeper sensitivities (at the level of tens of $\mu$Jy or below), and sub-arcsecond spatial resolutions are already providing many polarization detections (see \cite{heald2020} and references therein). These can contribute to polarimetric spectral characterizations of the AGN population (even towards its fainter ends), which could significantly differ from the total intensity one \cite{Galluzzi2017,Galluzzi2018}, and provide insights about the intrinsic geometry and structural complexity of emitting objects.

The radio spectra of SFGs are a combination of rather flat ($\alpha \sim-0.1$) free-free components from HII regions containing young massive ionizing stars and steeper ($\alpha \sim -0.7$) synchrotron components resulting from relativistic electrons accelerated by supernova remnants. On top of that, an additional synchrotron emission from a small-scale jet or winds/outflows associated with the nuclear activity may be present with various (and variable) spectral behaviors. As frequency increases, i.e., in mm or sub-mm regimes, the radio emission is progressively overwhelmed by the rise of the grey-body component due to dust emission associated with star formation. In fact, dust grains that are heated by the intense radiation from newly formed stars are not perfect absorbers, and the resulting emitting spectrum is determined not only by the temperature (as it happens in the black-body spectrum) but depends also on the opacity \cite{Calzetti2001}. Therefore, the usual SFG/RQAGN dichotomy is far from being sharp, and the heterogeneity of the RQAGN population is quite broad. Moreover, as the concept of radio quietness is inherently tied to observational constraints (comparing optical and radio domains), numerous sources once labeled as radio-quiet are now found to produce a weak yet detectable radio signal. This appearance of radio emission may result from nascent jets and, potentially, winds and outflows, serving as possible indicators of recently activated AGN. The co-eval improvement of facilities across the radio-optical bands strengthens the expectation of more and more refined classification according to radio-loudness.

In this paper, we will present the updated version of the source counts compilation by \cite{dezotti2010}, linking the old website (still accessible {{\url{http://w1.ira.inaf.it/rstools/srccnt/srccnt_tables.html}}}, accessed on 28 March 2025) 
 to a newly upgraded repository (see Section \ref{sec:repo}). Although a comprehensive review of source counts is beyond our present scope, we direct the reader to \cite{dezotti2010} for a summary of the fundamental ingredients and techniques.
 To complement the repository description, we will briefly summarize the new available technologies and the current status of radio surveys (see Section \ref{sec:surveys}) able to provide detections at sub-mJy fluxes, mostly focusing on the $\lesssim 2$ GHz range where, thanks to \texttt{SKA} precursors and pathfinders, most of the recent developments have arisen.

Furthermore, towards the end of this decade and during the next, new facilities like the \texttt{SKA} and the \texttt{ngVLA} are going to push current knowledge down to the $\mu$Jy regime in total intensity, unveiling polarimetric properties of SFGs and RQAGNs with total intensity fluxes as faint as few hundreds of $\mu$Jy. In light of this, we will briefly discuss issues that should be accounted for in building the counts distribution (see Section \ref{sec:issues}) and review useful methods for source counts derivations, e.g., for source counts convolution with polarization fraction distributions, source counts interpolation or extrapolations (see Section \ref{sec:methods}).

Finally, we will summarize the current status and future perspectives of radio source counts (see Section \ref{sec:conclusions}). 

\section{Recent 
 Radio Surveys and New Available Technologies}\label{sec:surveys}

New correlators with broader bandwidths have been one of the most significant technical improvements of radio interferometers in the last decade (e.g., \cite{wilson2011,morrison23}).
A factor $n$ in the bandwidth corresponds to a reduction by a factor of $\sqrt{n}$ in the observing time, assuming that all the other conditions have remained the same. For instance, the \texttt{VLA} (now called K. Jansky \texttt{VLA} or \texttt{JVLA}, and evolving towards the next-generation \texttt{VLA}, \texttt{ngVLA}, \cite{selina23})  improved its bandwidth size by a factor $\sim$80, and the Australia Telescope Compact Array (\texttt{ATCA}) of a factor $16$, thanks to the Compact Array Broadband Backend (\texttt{CABB} {\cite{wilson2011}}).

Thanks to further technological advances, broadband backend upgrades (dubbed as ultra-wide backends) have been still going on in these years and the next decade. Evidence of this revolution is the transition from \texttt{CABB} to \texttt{BIGCAT} for \texttt{ATCA} and the Wideband Sensitivity Upgrade (\texttt{WSU}) for \texttt{ALMA} {\cite{Gonzales2024}}. Thanks to these changes, \texttt{ATCA} will double its bandwidth, and \texttt{ALMA} will improve it by a factor $\sim$2--4. This, together with other planned upgrades in the receiver’s performance and signal path, will enhance the instantaneous telescope sensitivity of \texttt{ALMA} by a factor $\sim$2. The enhanced bandwidth for survey telescopes can be exploited either to increase sensitivity in a given region or to enhance the survey speed, allowing the covering of larger areas in a given time.

In this framework, there is another technical achievement that has been significantly improving the survey speed of modern radio facilities by at least an order of magnitude, i.e., the Phased Array Feed (\texttt{PAF}) technology, implemented, for example, on \texttt{ASKAP}, and under discussion for future upgrades of several facilities. A \texttt{PAF} is made of closely packed antenna elements (with spacing $\sim\lambda/2$) that, by spatially sampling the focal plane, can synthesize multiple independent beams electronically steerable (in real time and even reconfigurable at post-processing level), reaching Nyquist-sampling in the plane of the sky, in principle without requiring interleaved pointings (required instead by multi-feed receivers). Therefore, such a technological solution is flexible enough to perform very well with ultra-wideband receivers, as it can improve antenna efficiency and bandpass response over a very wide bandwidth, helping in mitigating radio frequency interference (RFI) and minimizing instrumental effects, such as off-axis aberration.

\begin{figure}[H]
    
    \includegraphics[width=0.75\linewidth]{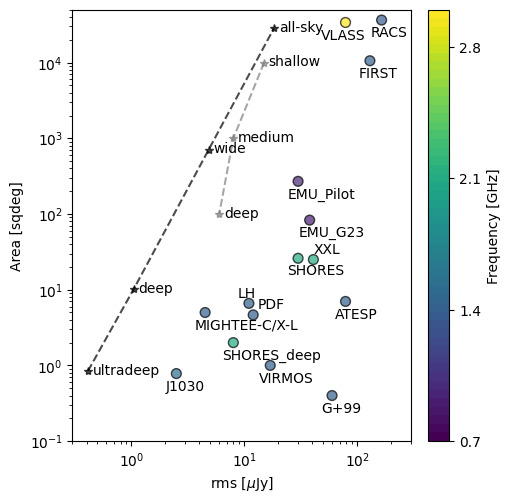}
\caption{Comparison 
 of sensitivity ($1\,\sigma$), surveyed area, and central observing frequency (color scale) for the available surveys reaching $1\sigma$ sensitivities $\lesssim 0.3\,$mJy in the frequency range $0.5\lesssim \nu\lesssim 3\,$GHz. {For 
 reference, we include the estimates for \texttt{SKA-MID} science cases by \cite{prandoni15} with the caveat that these possibilities are being revised while the instrument is developing but are still comparable with the feasibility of more recently planned science cases, defined for \texttt{SKA-LOW.} 
 \url{https://www.skao.int/sites/default/files/documents/d35-SKA-TEL-SKO-0000015-04_Science_UseCases-signed.pdf} (accessed on 28 March 2025).}
}
\label{fig:surveys_overview}
\end{figure}

Figure \ref{fig:surveys_overview} displays a sensitivity vs. covered area diagram for most of the surveys reaching the sub-mJy sensitivity range (note that some of them, like the \texttt{VLA} Sky Survey, \texttt{VLASS} {\cite{lacy20}}, Rapid \texttt{ASKAP} Continuum Survey, \texttt{RACS} {\cite{mcconnell20}}, and Evolutionary Map of the Universe, \texttt{EMU} {\cite{norris21}}, are still ongoing). This plot also reports for reference the estimates for some \texttt{SKA-MID} science cases by \cite{prandoni15} with the caveat that the possibilities are being revised while the instrument is being developed.

\section{Source-Count Repository}\label{sec:repo}

Differential source counts are defined as the number of sources per unit of area on the sky per flux-density bin, i.e., ${\rm d}N(S)/{\rm d}S$. Historically, they have been crucial in the debate between steady-state and Big-Bang cosmologies \cite{Ryle1955}. In the former Euclidean case, the number of uniformly distributed sources is proportional to the volume as $N\propto r^3$ for a sphere of radius $r$, while their flux-density scales as $S\propto r^{-2}$, so that $N \propto S^{-3/2}$, hence ${\rm d}N(S)/{\rm d}S \propto S^{-5/2}$. Thus, a normalization factor $S^{5/2}$ has often been employed to stress the discrepancy of the measured counts from the Euclidean case due to the expanding cosmological volume. That has been one of the first pieces of evidence in favor of the Big-Bang cosmology and has become the `typical' representation of the differential source counts ever since {\cite{ryle1961,longair1966,kellermann87,goss2023}}. 

\begin{figure}[H]

\includegraphics[width=\textwidth]{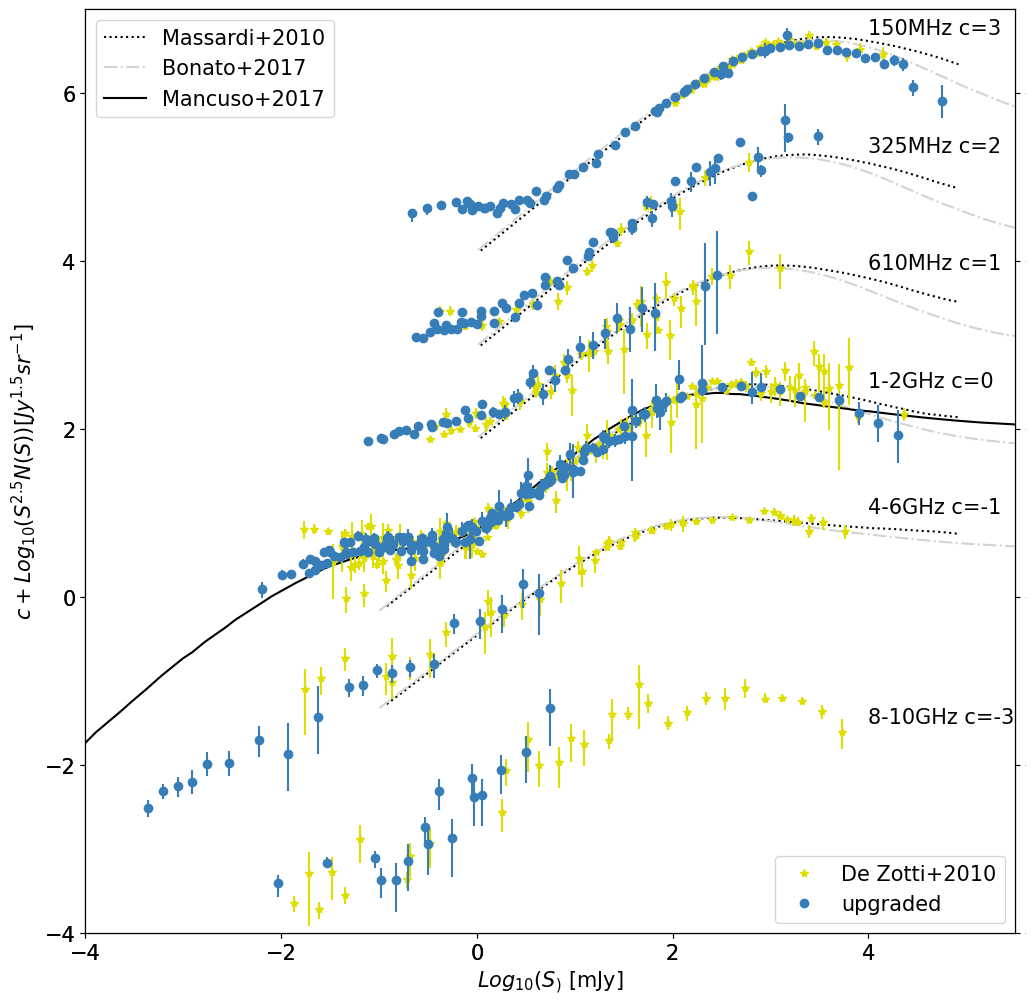}
\caption{Updated 
 radio source counts compilation at all the frequencies (blue dots) with respect to the version collected by \cite{dezotti2010} (yellow stars). Models accounting for sub-populations from \cite{MAssardi2010, Bonato2017,Mancuso2017} are also reported. Please note that to avoid overcrowding each frequency has been moved in the y-axis by a constant 'C' as declared in the plot labels. References for data are included in Table \ref{tab:counts_repo}.}
\label{fig:counts_all}
\end{figure}

Figure \ref{fig:counts_all} and Table \ref{tab:counts_repo} present all the upgrades of the new repository that can be found at the link 
 \url{https://sites.google.com/inaf.it/radio-source-counts} (accessed on 28 March 2025). We stress that this list includes the dataset uploaded to the repository at the time of writing the present paper: the repository will be monitored periodically in the future so as to be kept up to date as much as possible (see also \cite{tompkins23} for a recent summary of source counts data).

\begin{table}[H] 
\caption{The table summarizes the data that are collected in the updated repository, including some basic information of the original data like frequency, reference to the main paper, telescope, surveyed field(s), minimum flux density in log[Jy], and an indication of the overall area of the survey.\label{tab:counts_repo} An 'x' in the last column indicates if the dataset was already included in the \cite{dezotti2010} collection.}

\begin{adjustwidth}{-1.5cm}{0cm}
\centering 
\footnotesize
\begin{tabularx}{\fulllength}{LcCccCC}
\toprule
\textbf{Freq}	& \textbf{Ref}	& \textbf{Telescope}& \textbf{Survey, Field} &\boldmath{$\log S_{min}$}\textbf{[Jy]} & \textbf{Area [sqdeg]} & \textbf{In \cite{dezotti2010}}\\
\midrule
150 MHz &\cite{hales88}	& MRAO   &	6C, $>30\deg$ & $-$0.677 
 & 2030 & x\\
150 MHz &\cite{mcgilchrist90}	& CLFST   &	7C, 2fields & $-$0.978 & 472.7& x\\
150 MHz &\cite{williams16}	& LOFAR   &	Bootes & $-$3.149 & 19 &\\
150 MHz &\cite{franzen19}	& MWA   &	GLEAM, $<+30\deg$ & $-$1.161 & 24831&\\
150 MHz &\cite{mandal21}	& LOFAR   &	LoTSS,$>0\deg$  & $-$3.658 & 75.8&\\
\midrule
325 MHz &\cite{oort88}	& WSRT   &	Lynx  & $-$2.175 & 50 & x \\
325 MHz &\cite{owen09}	& VLA   &	SWIRE Field  & $-$3.372 & 1 & x\\
325 MHz &\cite{sirothia09}	& GMRT   &	ELAIS-N1  & $-$3.435 & 1.2 & \\
325 MHz &\cite{riseley16}	& GMRT   &	Super-CLASS  & $-$3.616 & 6.5&\\
325 MHz &\cite{mazumder20}	& GMRT   &	Lockman Hole  & $-$3.398 & 36 &\\
\midrule
610 MHz &\cite{katgert79}	& GMRT   &	4 fields  & $-$1.552 & 5 & x\\
610 MHz &\cite{bondi07}	& GMRT   &	VVDS-VLA  & $-$3.432 & 1 & x\\
610 MHz &\cite{moss07}	& GMRT   &	1H XMM-Newton/Chandra  &$-$3.310 & 64 & x\\
610 MHz &\cite{garn08}	& GMRT   &	 xFLS, ELAIS-
N1 and Lockman Hole  & -3.480 & 18 & x\\
610 MHz &\cite{whittam17}	& GMRT   &	 AMI001 field & $-$3.959 & 0.331 & \\
610 MHz &\cite{ocran20}	& GMRT   &	ELAIS-
N1  & $-$4.108 & 1.86 & \\
\midrule
1400 MHz &\cite{ciliegi99}	& VLA   &	ELAIS  & $-$3.770 & 4.22 & x \\
1400 MHz &\cite{gruppioni99}	& ATCA   &	ELAIS  & $-$3.569 & 4 & x \\
1400 MHz &\cite{richards00}	& VLA   &	HDF & $-$4.284 & 0.1 & x \\
1400 MHz &\cite{kellermann03}	& VLA   &	CDFS  & $-$4.206 & 4 & x \\
1400 MHz &\cite{hopkins03}	& ATCA   &	Phoenix  & $-$4.244 & 4.56 &x\\
1400 MHz &\cite{fomalont06}	& VLA   &	SSA 13 Field & $-$4.467 & 0.4 & x \\
1400 MHz &\cite{bondi08}	& VLA   &	COSMOS  & $-$4.180 & 2 & x \\
1400 MHz &\cite{owen08}	& VLA   &	Deep SWIRE  & $-$4.770 & 0.44 & x \\
1400 MHz &\cite{Seymour2008}	& VLA   & CDF & $-$4.340 & 0.196 & x \\
1400 MHz &\cite{Prandoni2018}& GMRT   & Lockman Hole & $-$4.036 & 6.6& \\
1400 MHz &\cite{Heywood2020}& VLA & XMM$-$LSS/VIDEO & $-$3.957 & 5 &  \\
1400 MHz &\cite{Matthews2021}& MeerKAT & DEEP2 & $-$4.900 & 1.04 &  \\
1400 MHz &\cite{DAmato2022}& VLA   & J1030 & $-$4.656 & 0.2 &  \\

2100 MHz &\cite{Butler2018}	& ATCA   & XXL & $-$3.550 & 25 &  \\
2100 MHz &\cite{Massardi2025}& ATCA   & SHORES & $-$3.383 & 26 &  \\
3000 MHz  &\cite{VanderVlugt2021}	& VLA   & COSMOS & $-$5.197 & 0.5 &  \\ 
3000 MHz &\cite{smolcic17}	& VLA   & COSMOS & $-$4.715 & 2 &  \\
\midrule
4760 MHz &\cite{altshurer86}	& GB   & $\delta=33\deg$ & $-$1.785 & 131 &  x \\
4850 MHz &\cite{donnelly87}	& VLA   & Linx.2 & $-$3.928 & 0.13 &  x \\
4850 MHz &\cite{gregory96}	& GB & GB6$0\deg<\delta<75\deg$ & $-$1.668 & 11003 & x\\
5000 MHz &\cite{fomalont84}	& VLA & multi fields & $-$3.862 & 0.004 & x \\
5000 MHz &\cite{fomalont91}	& VLA  & 1 field & $-$4.750 & 0.05 & x \\
5000 MHz &\cite{kuehr81}	 & VLA  & NRAOMPl+Parkes2.7GHz & 0.032 & 32207 & x \\
4850 MHz &\cite{pauliny80}	& VLA  & multi fields  & $-$1.921 & 23.4 & x \\
5000 MHz &\cite{wrobel90}	& VLA  & multi fields  & $-$2.721 & 2.27 & x \\
5500 MHz &\cite{huynh15}	& ATCA  & CDFS  & $-$4.310 & 0.34 &  \\
5500 MHz &\cite{prandoni06}	& ATCA  & ATESP  & $-$6.357 & 1 &  \\
\midrule
8500 MHz &\cite{fomalont02}	& VLA & multi fields & $-$4.863 & 653  & x \\
8500 MHz &\cite{henkel05}	& VLA & multi fields & $-$3.678 & 1.4 & x \\
8440 MHz &\cite{windhorst93}	& VLA & 2 fields & $-$4.717 & 0.027 & x \\
9000 MHz &\cite{huynh20}	& ATCA & CDFS & $-$3.975 & 0.276  &  \\
10000 MHz &\cite{jimenez24}	& VLA & GOODS-N &  $-$5.031 & 0.08 & \\
\bottomrule
\end{tabularx}
\end{adjustwidth}
\end{table}

\begin{figure}[H]

\includegraphics[width=\textwidth]{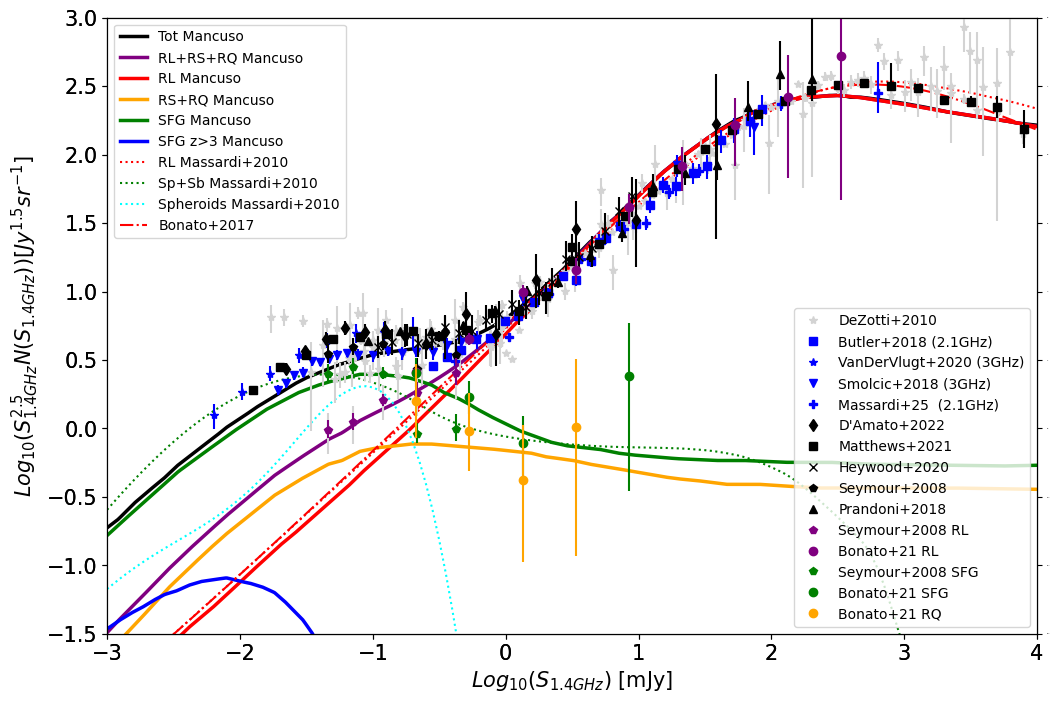}
\caption{Updated radio source counts compilation. Data are from \cite{dezotti2010,Butler2018,VanderVlugt2021,Smolcic2018,Massardi2025,DAmato2022,Matthews2021,Heywood2020,Seymour2008,Prandoni2018,Bonato2021}. Models accounting for sub-populations from \cite{MAssardi2010,Mancuso2017} are also reported.}
\label{fig:literature_source_counts}
\end{figure}

\textls[-15]{Figure \ref{fig:literature_source_counts} presents a collection of recent $\sim$2 GHz differential source counts \cite{dezotti2010,Butler2018,VanderVlugt2021,Smolcic2018,Massardi2025,DAmato2022,Matthews2021,Heywood2020,Seymour2008,Prandoni2018,Bonato2021, Bonato2021b}}, derived from the aforementioned surveys or a compilation of heterogeneous smaller ones. The figure also reports models that account for sub-populations, including the fainter ones \cite{MAssardi2010,Mancuso2017}. We remark that to compute source counts at a given frequency/wavelength, the reliability and completeness (down to a given flux-density level {\cite{jauncey75}}) of a source catalog are fundamental requirements to avoid systematic effects and inaccuracies. Those translate into great and usually unfeasible, at present, observational efforts, e.g., for multi-frequency coverage and source cross-band identifications. This explains why several of the deepest and/or widest area source counts are derived from the assembly of different smaller surveys, with at least a consistent overlap in frequency, flux density, and/or spatial coverage.

Integral counts $N(>S)$ report how many sources per unit sky area are brighter than a given flux-density threshold as a function of such limit {\cite{condon88,longair1974}}. These provide intuitive indications about the abundance of sources for a given area in the sky and constitute useful information for realizing simulations or planning surveys (covered area, sensitivity, and desired spatial resolution, e.g., in order to avoid source confusion). In fact, their slope provides a hint about the more efficient observational strategy to follow in order to statistically characterize a given population; e.g., if the slope is steeper than $-2$, it is more efficient to go deeper on small areas, while, if the slope is flatter, it is more practical to cover a wider area but to a shallower flux-density level (see also \cite{dezotti2010} for further details). Despite this, integral source counts are of little use compared to differential source counts, as they smear rapid changes with flux density, and numbers in different flux-density bins are not independent by definition.

\section{Errors, Biases, Corrections and Estimates}\label{sec:issues}
Throughout all the steps from data calibration to source extractions, a number of effects may arise and combine, inducing uncertainties and systematics in the reliability of detections, flux-density errors, as well as losses of detectable sources {\cite{kellermann87,prandoni01,bondi08,heywood13,zwart15}}. As a result, the final source counts may be significantly biased, especially in the faint end.

Thanks to the improved calibration recipes (e.g., direction-dependent approaches {\cite{albert20,roth23}}) and very accurate flux-density models for calibrators, residual imperfections in phase or amplitude induce flux-density errors typically below $5\,\%$. Similarly, new deconvolution software (such as \textsc{wsclean} \cite{Offringa2014}, \textsc{DDFacet} \cite{Tasse2018} and \textsc{CASA tclean} \cite{CASA2021}) implements accurate telescope response, in terms of both synthesized and primary beam, also including improved algorithms for w-correction and multi-frequency synthesis: the latter allows minimizing the bandwidth smearing effect, which consists of a radially blurred image becoming worse moving away from the phase center. This effect is particularly significant when the fractional bandwidth $\Delta\nu/\nu_0$ (where $\nu_0$ is the central frequency of the observation) is larger. Another key ingredient of modern imaging software is represented by optimized cleaning algorithms, combined with advanced auto-masking and multi-scale features, able to better characterize local noise variations; this reduces over-cleaning, which would lead to misinterpretation of artifacts surrounding bright sources to be deconvolved as real sources, or under-cleaning, which would lead to flux-density underestimation for more extended sources.

Despite these advancements, there are other effects to be carefully taken into account in survey design and its interpretation, such as the resolution bias, the confusion level, and the Eddington bias {\cite{condon98,prandoni01,Matthews2021}}. Telescopes like \texttt{SKA} and its precursor and pathfinders can attain spatial resolution at the arcsec level (or even below). This will open a window down to $\mu\,$Jy level sources without suffering from source confusion issues, i.e., the degradation of sensitivity performances due to the blending of discrete sources. At the same time, as the spatial resolution improves, the fraction of resolved sources increases. Therefore, if the surface brightness of an object falls below the detection threshold, one could underestimate its flux density, biasing the final catalog and the derived source counts. This is the effect of the well-known resolution bias. Again, imaging software and modern source extraction tools can only partially mitigate the impact of this effect. Thus, observations, simulations, or semi-analytic methods can help determine the correction factors for source counts at the different frequency bins \cite{Prandoni2018}.

The Eddington bias arises from the interplay of the noise distribution and the underlying flux-density distribution of sources, characterized by a steep behavior, i.e., many more fainter sources than brighter ones. Thus, a number of faint sources close to the detection levels can be boosted by noise fluctuations above the threshold, introducing a skewness in the recovered flux-density distribution, hence boosting the faint tail of source counts. Non-Gaussianities in the noise structure are usually present in radio images because of residual RFI, together with instrumental effects or bright sources inducing sidelobes. The combination of these instances worsens the boosting effect. The most effective way to mitigate it is to resort to statistical methods based on simulations, in order to infer the entity of the skewness from the comparison of the synthetic `measured' fluxes compared to the injected `true' ones.

Even after accounting for the effects mentioned above, determining accurate source counts requires addressing incompleteness in source detection due to noise, confusion, or limitations in the detection algorithm. This is particularly important at the faint end of the source counts, where undetected sources are more common, and correcting for this incompleteness reduces the associated error bars. One statistical method to address this is the $P(D)$ approach {\cite{condon2012}}. This method calculates the probability of source detection, $P(D)$, as a function of source flux, morphology, and observing conditions, especially the level of confusion. It assumes that the observed $P(D)$ results from the combination of the probability distributions of source confusion and noise and that its variance is related to the combination of the independent variances of these two distributions. Although high-resolution capabilities in the \texttt{SKA} era lessen the impact of confusion, and thus the importance of $P(D)$ compared to earlier surveys, it remains relevant. Specifically, $P(D)$ is still crucial for characterizing and understanding the impact of cosmic variance \citep{Heywood2013}. Cosmic variance refers to statistical fluctuations in the observed source density compared to the average density across the Universe. These fluctuations arise from large-scale structures like galaxy clusters and cosmic voids, coupled with the fact that observations cover only a limited portion of the sky.  

\section{Radio Source Counts Derivation Methods}\label{sec:methods}

Before the advent of \texttt{SKA} precursor and pathfinders, the only way to provide a more comprehensive picture of source populations was to combine data from multiple surveys. In some cases, this implies dealing with partially overlapping smaller surveys characterized by different sensitivities, resolutions, and other instrumental effects (e.g., slightly different frequency setups or residual calibration issues), such as the NRAO VLA Sky Survey (\texttt{NVSS} {\cite{condon98}}) and the Faint Images of the Radio Sky at Twenty centimeters (\texttt{FIRST} \cite{white97}). As an example, prior to combining data to derive source counts, a number of steps are required to cross-match original catalogs, cross-calibrate for flux densities, classify sources (mostly using spectral indexes), and finally assess the completeness and reliability of the combined catalog by comparing detection rates and consistency of source properties among the combined surveys.

A common but still delicate practice is extrapolating source counts to lower flux densities since it requires careful modeling of source-count distribution at faint fluxes and assumptions about potential biases, such as effects of source evolution, confusion, and other instrumental limitations. In this respect, model fitting combined with Markov Chain Monte Carlo (MCMC) methods (naturally suited for Bayesian inference) allows the incorporation of prior information about source populations and provides a natural framework for quantifying uncertainties associated with model parameters and, in turn, assessing uncertainties on extrapolated source counts.

Another common practice is to extrapolate (or, whenever possible, interpolate) source counts at different frequencies. 
This requires assumptions about the spectral energy distributions (SEDs) of various radio-source populations. 
Since spectral index distributions vary depending on the source type, it is crucial to consider redshift evolution when dealing with lower frequencies, where high-redshift sources contribute more significantly to the observed population.  

At higher frequencies, additional complications arise. Although source confusion decreases as a result of improved resolution, variability becomes a major concern. 
Flat-spectrum sources, including blazars and compact AGNs, often exhibit significant variability on short timescales, introducing uncertainties at both the low- and high-flux-density ends of the extrapolated source counts. 
These factors make it challenging to produce reliable models for high-frequency radio populations solely based on lower-frequency observations.

\begin{figure}[H]
    
    \includegraphics[width=0.75\linewidth]{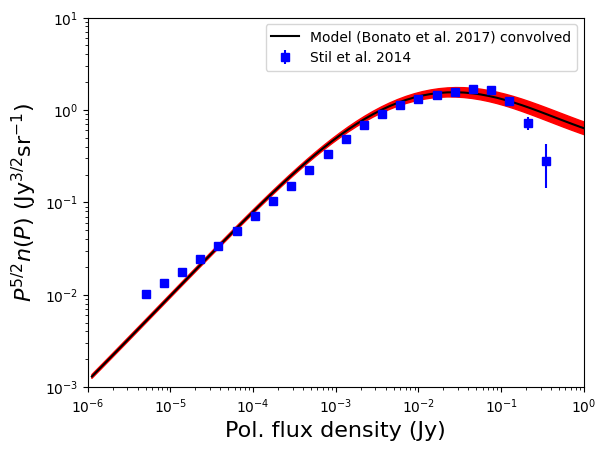}
\caption{Comparison 
 of polarization differential source counts from the model of \cite{Bonato2017} convolved with polarization fractions observed by \cite{Galluzzi2018}, with those estimated by \cite{Stil2014}. The red band displays the resulting $1\,\sigma$ uncertainty from observed polarization fractions of \cite{Galluzzi2018}.}
\label{fig:polarized_sourcecounts}
\end{figure}

As mentioned in Section \ref{sec:intro}, polarization fractions are usually low, related observations are time-consuming, and calibration strategies are more complex and difficult with respect to total intensity. Thus, deep full-Stokes' surveys complete down to sub-$\mu$Jy level would be needed in order to characterize the polarimetry of faint radio populations, and these are still missing. A possibility is to resort to stacking techniques: \cite{Stil2014} exploited \texttt{NVSS} to statistically derive a relationship between the total intensity flux density and the linearly polarized intensity by considering Stokes' Q and U separately and addressing debiasing via Monte Carlo simulations. They found a slightly increasing trend for the median polarization fraction up to $2-3\%$ at fainter total intensity fluxes ($\sim 1\,$mJy), and they did not attempt to select different populations. Another possibility to predict polarized radio source counts that can work for different populations independently but, again, implies some level of extrapolation is to convolve total intensity differential source counts with a distribution of polarization fractions, which can be simply assumed (from hints available from past studies) or measured from brighter objects of a given class. Therefore, such a method requires careful consideration of the sensitivity of polarimetric surveys, potential biases in polarization measurements, and assumptions about the relationship between total intensity and polarized emission for different source populations. Such a convolution can be represented as 
\begin{equation}
n(P)=\int_{S_0=P}^\infty {\mathcal{P}\left(m=\frac{P}{S}\right)n(S)\frac{dS}{S}}\; ,
\label{equ:difsoucouP}
\end{equation}
where $\mathcal{P}$ is the probability density distribution for the
polarization fraction $m=\Pi/100$. The integration over $S$ is truncated at $S_0=P$, where the polarization fraction
is $100\%$; however, since the distribution of observed polarization fractions is usually fitted with a lognormal distribution (which goes rapidly to zero for $\Pi> 10\%$; see \cite{Massardi2013,Galluzzi2017}), the result is insensitive to the choice of $S_0$ provided that it is not much larger than $P$. As an example of this approach,  Figure \ref{fig:polarized_sourcecounts} reports the convolution of the model for total intensity differential source counts (all populations) reported by \cite{Bonato2017} at $1.4\,$ GHz with a lognormal distribution derived from observed polarization fractions of \cite{Galluzzi2018} over the 1.1--3.1 GHz frequency range. The latter observations (encompassing $107$ objects drawn from the faint \texttt{Planck}-\texttt{ATCA} Co-eval Observations \texttt{PACO} sample; see \cite{Bonavera2011}) went as deep as $\sigma_P\sim 0.2$ mJy/beam. Given the relative paucity of the sample, we performed a bootstrap and resampling experiment with $10^3$ repetitions, fitting each synthetic sample with a lognormal distribution. The final set of parameters was derived as median values, and the resulting $1\,\sigma$ uncertainty in the convolved source counts is shown as a red band in the Figure. For comparison, polarized source counts by \cite{Stil2014} are also plotted as blue squares.      

\unskip 

\section{Summary and Future Perspectives}
\label{sec:conclusions}

In this work, we have compiled an updated repository of sub-mJy extragalactic radio source counts that can be found at the link 
\url{https://sites.google.com/inaf.it/radio-source-counts} (accessed on 28 March 2025), significantly enhancing our understanding of faint radio-source populations. The literature reveals the complexity and diversity of sources, such as Dusty Star-Forming Galaxies (SFGs) and Radio-Quiet AGNs (RQAGNs), emphasizing the importance of accounting for observational biases and systematic errors in source-count derivations. Despite the fact that dusty SFGs constitute the bulk of sub-mJy radio populations, they remain substantially unexplored due to the much shallower sensitivity limits of current radio surveys. At the same time, the origin of radio emission in RQAGNs remains debated, with proposed mechanisms including small-scale jets, shock fronts associated with AGN-driven outflows, winds originating from the outer regions of the accretion disk, and electron acceleration via magnetic reconnection in the thin-disk corona \cite{Panessa2019}.

The integration of advanced techniques, including polarization studies and enhanced sensitivity from next-generation radio telescopes like the Square-Kilometer Array (\texttt{SKA}), promises to further refine our knowledge of faint radio populations. In fact, the situation has been steadily improving thanks to the profusion of surveys planned for the coming years with \texttt{SKA} pathfinders and precursors (\texttt{EMU}, \texttt{MIGHTEE}, \texttt{RACS}, \texttt{VLASS}, etc.). These advancements will provide unprecedented insights into the evolution of galaxies and AGNs across cosmic times.

A crucial step forward will be the combination of deep radio observations with ancillary data from upcoming multi-wavelength facilities. Optical and near-infrared surveys from the \texttt{Vera C. Rubin Observatory}’s Legacy Survey of Space and Time (\texttt{LSST}) will provide precise photometric redshifts and morphological classifications essential for associating radio sources with their host galaxies. Meanwhile, next-generation X-ray observatories such as \texttt{NewAthena} 
 ({{\url{https://www.cosmos.esa.int/web/athena}}}, accessed on 28 March 2025) will help distinguish between AGN-driven radio emission and star formation processes by identifying obscured accretion activity in RQAGNs. At higher energies, the Cherenkov Telescope Array \texttt{CTA} 
 ({{\url{https://www.ctao.org}}}, accessed on 28 March 2025) will offer insights into the non-thermal emission of radio-loud AGNs, particularly in the context of jet physics and particle acceleration mechanisms. Combining these multi-wavelength datasets will be key to disentangling the nature of sub-mJy radio sources and constraining their evolutionary pathways.

Future work will focus on leveraging deeper surveys and expanded frequency coverage, enabling the exploration of even fainter flux regimes. As observational capabilities continue to grow, we anticipate substantial progress in characterizing the physical properties of the faintest radio sources, thereby enriching our understanding of galaxy evolution and the cosmic radio sky.
Overall, our repository serves as a foundational reference for upcoming radio surveys and a stepping stone toward a more comprehensive view of the radio Universe.




\vspace{6pt}
\authorcontributions{Conceptualization, V.G., M.B., M.G., A.L.; methodology, V.G.; software, V.G.; validation, M.B., M.G.; data curation, V.G.; writing V.G., M.B., M.G., A.L. All authors have read and agreed to the published version of the manuscript.}

\funding{This 
 work was partially funded from the projects: `Data Science Methods for MultiMessenger Astrophysics \& Multi-Survey Cosmology' funded by the Italian Ministry of University and Research, Programmazione Triennale 2021/2023 (DM n.2503 dd. 9 December 2019), Programma Congiunto Scuole; EU H2020-MSCA-ITN-2019 n. 860744 \textit{BiD4BESt: Big Data applications for black hole Evolution STudies}; Italian Research Center on High-Performance Computing Big Data and Quantum Computing (ICSC), project funded by European Union---NextGenerationEU---and National Recovery and Resilience Plan (NRRP)---Mission 4 Component 2 within the activities of Spoke 3 (Astrophysics and Cosmos Observations);  European Union---NextGenerationEU under the PRIN MUR 2022 project n. 20224JR28W ``Charting unexplored avenues in Dark Matter''; INAF Large Grant 2022 funding scheme with the project ``MeerKAT and LOFAR Team up: a Unique Radio Window on Galaxy/AGN co-Evolution; INAF GO-GTO Normal 2023 funding scheme with the project ''Serendipitous H-ATLAS-fields Observations of Radio Extragalactic Sources (SHORES)''.}

\dataavailability{The data presented in this study are openly available at the link \url{https://sites.google.com/inaf.it/radio-source-counts} (accessed on 28 March 2025).
} 

\acknowledgments{
We thank M. Massardi for useful discussions. We acknowledge the anonymous referees for constructive comments and suggestions.}

\conflictsofinterest{The authors declare no conflicts of interest.}

\begin{adjustwidth}{}{0cm}

\reftitle{References}

\PublishersNote{}

\end{adjustwidth}

\end{document}